# Memristive switching in the surface of a charge-density-wave topological semimetal

Jianwen Ma[1#], Xianghao Meng[2#], Binhua Zhang[3,4#], Yuxiang Wang[1], Yicheng Mou[1], Wenting Lin[5], Yannan Dai[6,7], Luqiu Chen[6,7], Haonan Wang[6], Haoqi Wu[8], Jiaming Gu[1], Jiayu Wang[1], Yuhan Du[2], Chunsen Liu[9], Wu Shi[1,10], Zhenzhong Yang[6], Bobo Tian[6,7], Lin Miao[5], Peng Zhou[8,9], Chun-Gang Duan[6,7], Changsong Xu[3,4*], Xiang Yuan[2,7,11*], Cheng Zhang[1,10*]

[1] State Key Laboratory of Surface Physics and Institute for Nanoelectronic Devices and Quantum Computing, Fudan University, Shanghai 200433, China

[2] State Key Laboratory of Precision Spectroscopy, East China Normal University, Shanghai 200241, China

[3] Key Laboratory of Computational Physical Sciences (Ministry of Education), Institute of Computational Physical Sciences, State Key Laboratory of Surface Physics, and Department of Physics, Fudan University, Shanghai 200433, China

[4] Shanghai Qi Zhi Institute, Shanghai 200030, China

[5] School of Physics, Southeast University, Nanjing 211189, China

[6] Key Laboratory of Polar Materials and Devices (Ministry of Education), Department of Electronics, East China Normal University, Shanghai 200241, China

[7] Shanghai Center of Brain-inspired Intelligent Materials and Devices, East China Normal University, Shanghai 200241, China

[8] State Key Laboratory of ASIC and System, School of Microelectronics, Fudan University, Shanghai 200433, China

[9] Frontier Institute of Chip and System, Fudan University, Shanghai 200433, China

[10] Zhangjiang Fudan International Innovation Center, Fudan University, Shanghai 201210, China

[11] School of Physics and Electronic Science, East China Normal University, Shanghai 200241, China

[#] These authors contributed equally to this work

[*] Correspondence and requests for materials should be addressed to C. X. (E-mail: csxu@fudan.edu.cn), X. Y. (E-mail: xyuan@lps.ecnu.edu.cn) & C. Z. (E-mail: zhangcheng@fudan.edu.cn)

**Abstract**

Owing to the outstanding properties provided by nontrivial band topology, topological phases of matter are considered as a promising platform towards low-dissipation electronics, efficient spin–charge conversion, and topological quantum computation. Achieving ferroelectricity in topological materials enables the non-volatile control of the quantum states, which could greatly facilitate topological electronic research. However, ferroelectricity is generally incompatible with systems featuring metallicity due to the screening effect of free carriers. In this study, we report the observation of memristive switching based on the ferroelectric surface state of a topological semimetal $(TaSe_4)_2I$. We find that the surface state of $(TaSe_4)_2I$ presents out-of-plane ferroelectric polarization due to surface reconstruction. With the combination of ferroelectric surface and charge-density-wave-gapped bulk states, an electric-switchable barrier height can be achieved in $(TaSe_4)_2I$–metal contact. By employing a multi-terminal-grounding design, we manage to construct a prototype ferroelectric memristor based on $(TaSe_4)_2I$ with on/off ratio up to $10^3$, endurance over $10^3$ cycles, and good retention characteristics. The origin of the ferroelectric surface state is further investigated by first-principles calculations, which reveals an interplay between ferroelectricity and band topology. The emergence of ferroelectricity in $(TaSe_4)_2I$ not only demonstrates it as a rare but essential case of ferroelectric topological materials, but also opens new routes towards the implementation of topological materials in functional electronic devices.

**MAIN TEXT**
**Introduction**

The developments in topological phases of matter have brought revolutionary changes to the understanding of electronic structure in condensed matter physics.[1–4] Characterized by topological invariants in the wave functions, topological materials present gapless surface/edge states as a result of bulk–boundary correspondence.[1] These topologically protected states are generally robust against disorder and give rise to exotic physical phenomena such as quantum (spin/anomalous) Hall effect[5–7], spin-polarized Dirac cone[8,9], and Fermi arc[10–13]. Intriguing transport properties[14–22], including ultrahigh mobility, giant magnetoresistance, and efficient spin-charge conversion, have been demonstrated in topological materials. The surging advances in the research of topological physics initiate broad prospects of utilizing topological materials in various electronic device applications[23–28], ranging from low-dissipation electronics and spintronics to quantum computation.

Coupling electric charges with other degrees of freedom, such as spin or lattice, could help to realize a non-volatile control of the quantum states. Magnetic topological insulators and semimetals are typical examples of correlating charges with spin. Recent progresses in magnetic topological materials have evoked a brand-new direction of topological spintronics and magnetoelectronics[25,26,29], which is now becoming a large branch of this field. On the other hand, electronic modulation through lattice is generally related to ferroelectricity or piezoelectricity. They are usually incompatible with metallic systems due to the screening effect of free carriers. Hence, achieving strong electron–lattice coupling, such as ferroelectricity should be potentially helpful for topological electronic research. So far, ferroelectric topological materials have only been found in limited cases[30–33], while related device research is rarely explored.

In this work, we report the observation of memristive switching based on the ferroelectric surface state of a charge-density-wave topological semimetal (TaSe$_4$)$_2$I. The presence of surface ferroelectricity in a topological system with bulk charge density wave gap enables an efficient electric switching of Schottky barrier height. We successfully construct the first topological-semimetal-based prototype ferroelectric memristor with on/off ratio up to $10^3$, endurance over $10^3$ cycles, and good retention characteristics. Our results establish (TaSe$_4$)$_2$I as a rare but essential case of ferroelectric topological materials and open up new directions for novel topological electronic devices.

**Results**

(TaSe$_4$)$_2$I is a quasi-1D material with a body-centered tetragonal unit cell (space group I422, no. 97) at room temperature.[34] Each tetragonal unit cell is composed of TaSe$_4$ chains, which are parallelly aligned along the *c*-axis, and separated by strands of iodine atoms (Fig. 1a). The room-temperature phase of (TaSe$_4$)$_2$I is proposed to be a Weyl semimetal with 24 pairs of Weyl points.[35,36] A charge density wave transition at $T_c$ = 263 K results in a gap opening of about 0.2 eV [37–41], providing a route towards axion electrodynamics in topological materials.[35,42] Evidence of possible axionic charge density wave has been reported in previous transport experiments.[43] In this work, needle-like single crystals of (TaSe$_4$)$_2$I (Fig. S1a) were synthesized by the chemical vapor transport method (refer to Methods). The as-grown (TaSe$_4$)$_2$I crystals are characterized by Raman spectra (Fig. 1b). Similar to previous reports, three dominating peaks at 146, 183, and 272 cm$^{-1}$ as marked by the dashed lines correspond to the $A_1^1$, $A_1^2$, and $A_1^3$ modes, respectively.[44] Fig. 1c presents the temperature dependence of resistivity $\rho$ with electric current applied along the *c* axis of the crystals. The monotonic decrease of $\rho$ with the increase of *T* suggests strong thermal activation with an activation energy of 137.2 meV (the inset of Fig. 1c). The kink around 263 K corresponds to the charge density wave transition. Below the transition temperature, a gap will be generated by the forming of density wave, resulting in the thermal activation behavior. Fig. 1d is the X-ray photoelectron spectroscopy (XPS) of (TaSe$_4$)$_2$I crystals, in which the chemical states of Se, Ta, and I atoms can be identified. We further measured the electronic structure of (TaSe$_4$)$_2$I using angle-resolved photoemission spectroscopy (ARPES). As shown in Fig. 1e, two sets of Dirac-like energy bands are observed along the $\overline{Z\Gamma Z}$ direction in both the ARPES intensity (left panel) and the corresponding second-order differential band map (right panel) at room temperature. The constant energy contour (CEC) mapping acquired on the $\overline{X\Gamma Z}$ plane at $E$-$E_F$ = -0.1 eV (Fig. 1f) indicates strong electronic anisotropy from the quasi-1D crystal structure. The obtained electronics structure is consistent with the presence of topological semimetal state in (TaSe$_4$)$_2$I proposed earlier[35,36,45].

Taking advantage of the weak bonding between TaSe$_4$ chains, we obtain (TaSe$_4$)$_2$I nanoribbons from bulk crystals through mechanical exfoliation. The typical thickness of exfoliated nanoribbons used in this study is around a few hundred nanometers (refer to Fig. S1b for the atomic force microscope image of a 300-nm nanoribbon). We start by fabricating a (TaSe$_4$)$_2$I nanodevice with tri-layer Cr/Ag/Au electrodes. The resistivity at different experimental configurations is labeled as $R_{i-j}^{m-n}$ where $i,j$ and $m,n$ denote the source probe and the detect probe, respectively. Specifically, the grounded probe is denoted by $G$. The left panel of Fig. 2a exhibits the total resistance $R_{3-G}^{3-G}$ measured from Electrode 3 at 120 K while all remaining electrodes are grounded. As the bias voltage is swept in a sequence of 0 V→1 V→0 V→-1 V→0 V, a clear resistance transition between a high-resistance state (HRS) and a low-resistance state (LRS) with hysteresis can be observed, with a sizeable on-off ratio of nearly three orders of magnitude. It corresponds to a typical bipolar switching property as a memristor[46–49].

To investigate the origin of the memristive behavior, we compare other two measurement configurations as illustrated in the middle and right panels of Fig. 2a. When applying the bias voltage at Electrode 3 and grounding 1, 4, and 5, we measure the voltage $U$ between 3 and 2 as well as the current $I$ through 3. It corresponds to the resistance $R_{3-G}^{3-2}$ involving the contact resistance at Electrode 3 as well as the channel resistance between 3 and 2. Similar bipolar switching also appears in $R_{3-G}^{3-2}$. In contrast, when passing current through 1 and grounding 4 and 5, the resistance measured between 2 and 3 ( $R_{1-G}^{2-3}$ ) shows a much smaller value (around 1.1 kΩ) without the bias-voltage-dependent switching behavior. Here, $R_{1-G}^{2-3}$ corresponds to the channel resistance of (TaSe$_4$)$_2$I between 2 and 3, which is over three orders of magnitude smaller than the peak resistance value of HRS. The comparison of the three panels in Fig. 2a suggests that the contact resistance between metal electrodes and (TaSe$_4$)$_2$I channel is responsible for the bipolar switching property in $R_{3-G}^{3-G}$. Similar memristive behavior can also be detected in bulk (TaSe$_4$)$_2$I samples (shown in Fig. S2). Meanwhile, in Fig. S3, we track the evolution of Raman spectrum in bulk (TaSe$_4$)$_2$I channel during the resistance switching process when sweeping external bias voltage. No significant change in Raman spectrum is observed as the resistance is tuned to HRS at large biases, which excludes the possibility of bias-induced structure phase transition in the inner channel between electrodes.

Having established the link between the contact resistance and the memristive behavior, we now turn to investigate the surface property of (TaSe$_4$)$_2$I and the underlying mechanism. The presence of ferroelectricity in (TaSe$_4$)$_2$I is demonstrated by piezoresponse force microscopy (PFM). As a standard characterization technique for ferroelectric and piezoelectric materials, PFM detects the dynamics of electric polarization in external electric fields through the electromechanical interaction between the sample and tip. The inset of Fig. 2b shows the schematics of room-temperature PFM measurement on a (TaSe$_4$)$_2$I nanoribbon exfoliated on a conductive silicon substrate. A bias voltage between -30~30 V is applied between the PFM tip and silicon substrate while monitoring the local PFM phase and amplitude. Fig. 2b and c are the PFM phase and amplitude loops versus the bias voltage at room temperature, respectively. During the sweeping process of bias voltage, the PFM amplitude presents a typical butterfly shape, accompanied with a phase change of approximately 180° at the minima of amplitude. The asymmetry in PFM amplitude loop is likely to originate from the typical electrostatic effect during the experiment[50,51].These results serve as clear evidences for the presence of out-of-plane ferroelectric polarization in (TaSe$_4$)$_2$I under external electric fields.

Since the bulk crystal structure of (TaSe$_4$)$_2$I belongs to a non-polar point group[34,35,39], which does not support the ferroelectricity, it is reasonable to consider the surface origin of the observed ferroelectric behavior. Before turning to discuss the physical origin of ferroelectricity in (TaSe$_4$)$_2$I, we first discuss the mechanism of the memristive behavior formed in the metal–(TaSe$_4$)$_2$I junction. Since (TaSe$_4$)$_2$I is in the charge-density-wave phase with large bulk resistivity at low temperatures, a Schottky barrier can be formed in the metal contact regions. The out-of-plane ferroelectric polarization (left or right polarization as schematized in Fig. 2d–e) will affect the Schottky barrier depending on their relative orientations. When a negative voltage (above the threshold field for ferroelectric switching) is applied to the metal–(TaSe$_4$)$_2$I junction (Fig. 2d), the surface ferroelectric polarization will be switched towards the metal contact, resulting in a downward shift of the band bending near the interface. In this case, it is easier for electrons to transport across the junction, which gives the LRS. Here, the insulating bulk state as well as the confined surface state help to suppress the carrier screening along the out-of-plane direction, which allows an effective switching of ferroelectric polarization by bias voltage. Conversely, a positive voltage above the threshold at the junction leads to ferroelectric polarization from the metal side to the $(TaSe_4)_2I$ side (Fig. 2e). The rapid increase of Schottky barrier height gives the HRS. In this way, memristive behavior based on ferroelectric switching will be induced when scanning the bias voltage back and forth. We also note that in additional to the Schottky barrier, a surface–bulk barrier may also appear due to the accumulation or depletion of charges near the surface. Nevertheless, its presence only affects the detailed band bending profile shown in Fig. 2d (a peak- or dip-like feature may appear in the middle due to surface charge accumulation or depletion). And the surface–bulk barrier should be smaller than that between metal and $(TaSe_4)_2I$ surface. Therefore, the barrier modulation mechanism due to ferroelectric polarization remains valid.

  Back to the device used in this study, a multi-terminal grounding configuration is employed as illustrated in the inset of Fig. 2a. When passing a proper current through Electrode 3 in the middle while grounding the rest four electrodes, the bias electric field at Electrode 3 can exceed the threshold value for the ferroelectric polarization switching. During the *I*–*V* scan, other contact regions are kept un-switched due to the current shunting effect (so that the corresponding electric fields are not strong enough to switch the ferroelectric polarization at other contact regions). Otherwise, for a conventional two-terminal geometry, the ferroelectric polarization for source and drain contacts will be switched to the opposite direction simultaneously, resulting in the coexistence of a LRS and a HRS throughout the voltage scanning process (refer to Fig. S5). In other words, the multi-terminal grounding configuration breaks the symmetry between source and drain contacts, enabling the emergence of the memristive property. The presented model can well explain the obtained results in Fig. 2a.

  We now turn to characterize the device performance of $(TaSe_4)_2I$–metal memristor devices. The following tests of memristive performances in Figs. 3–4 were measured in the multi-terminal grounding configuration. Fig. 3a presents the current-voltage (*I*–*V*) characteristics in DC voltage sweep measurements with the corresponding device optical image shown in the inset. The experimental temperature is fixed at 120 K unless specified. The result is illustrated in semilogarithmic with the sweeping direction indicated by arrows. With the steady increase of forward voltage, a pronounced change of resistance from LRS to HRS is observed at around 0.5 V (the current value at LRS is 108.9 μA, corresponding to a current density of 279.2 μA/μm²), which is called the "RESET" process. Subsequently, when sweeping the voltage reversely to negative values, an opposite "SET" process takes place, which corresponds to the bipolar switching property. Fig. 3b presents the *I*–*V* characteristics at different temperatures from 40 K to 120 K with clear bipolar switching and good memristive behavior. The nanodevices also show good air stability at and below room temperature (refer to Fig. S6). The temperature dependence of the Set and Reset voltages ($V_{Set}$ and $V_{Reset}$) is summarized in Fig. 3c, showing a systematic decrease with the increase of temperature. By taking a typical depletion layer width in the order of 100 nm[52], we roughly estimate the effective field responsible for the resistance switching as 12.2 to 5.6 MV/m from 40 to 120 K. Despite the persistence of ferroelectricity up to room temperature, the switching behavior cannot be clearly resolved above 180 K. It may result from the dramatic decrease of Schottky junction resistance at elevated temperatures.

  The endurance test is carried out to examine the switching stability of the nanodevices at 120

K. Under repetitive cycles of programmed set → read → reset → read pulses (set pulse: -2.5 V, read pulse: 0.05 V, reset pulse: 2.5 V, pulse duration: 0.5 s), the device shows reproducible switching with repeatable LRS and HRS states for over 1000 times of cycling (Fig. 4a). Fig. 4b is the retention of LRS and HRS, which presents good stability over 10 hours, manifesting a good non-volatile property. We also investigate the required pulse number at different set voltages (0.4 V, 0.6 V, 0.8 V, and 1 V) to switch the resistance state as presented in Fig. 4c. It systematically decreases from 876 times at 0.4 V to 26 times at 0.5 V. Above 0.6 V, the resistance switching can be achieved with the application of one single pulse. Furthermore, distinct multiple levels of resistance state between HRS and LRS can be obtained by applying different pulse numbers with a set voltage of 0.4 V (Fig. 4d). The response time of two typical devices was measured as 1.69 and 0.67 ms, respectively (Fig. S7).

To address the device-to-device reproducibility, the distributions of $V_{Set}$ and $V_{Reset}$ are obtained by measuring 31 nanodevices. As shown in Fig. 4e, the set voltage is distributed in the range of -0.1 V to -0.9 V, with a peak voltage around -0.3 V. The reset voltage is distributed in the range of 0.2 V to 1 V with a peak voltage around 0.5 V. In addition, we calculate the resistance off/on ratio between LRS and HRS, $R_{off}/R_{on}$, for these 31 different devices, as shown in Fig. 4f. The variation of $R_{off}/R_{on}$ mainly results from different contact interface quality among devices, which affects the Schottky barrier height. In Fig. S2, we show that the similar memristive behavior can be detected in bulk $(TaSe_4)_2I$ devices with multi-grounding configuration. In contrast, the $R_{off}/R_{on}$ from the bulk devices significantly decreases compared to the value from nanodevices in Fig. 4f. It is likely to result from the large contact area in bulk samples, which may provide defect-induced conducting regions across the Schottky barrier and diminish the ferroelectric switching performance. Since a sizeable Schottky barrier is required for resistance switching, the memristive property in $(TaSe_4)_2I$ currently is limited to low temperatures. Further optimization may be carried out by increasing the transition temperature of charge density wave or using other types of barriers valid for the semimetal phase.

**Discussion**

Having established the ferroelectric memristive behavior, we now analyze the possible origin of the observed ferroelectricity in $(TaSe_4)_2I$. We performed density-functional theory (DFT) calculations on the experimentally observed $(TaSe_4)_2I$ (110) surface with I-terminated slab structures (Fig. 5a)[45]. By comparing various possible configurations that associate with ferroelectricity, we indicate that the surface reconstruction of I atoms is more favored due to its lower energy barriers and ferroelectric states (see Fig. S10, 11 for the ferroelectricity induced by the mobility of I atoms and the sliding of the $TaSe_4$ chain). This is understandable, since the cleaving process occurs at the I termination[45], the I atoms can easily slide from the cleaved surface, resulting in the surface reconstruction of I atoms in the outermost and second outermost surface layers (Fig. 5b). As a result, the I layer on surface buckles dramatically, effectively breaking the centrosymmetry and leading to the spontaneous out-of-plane electric polarization. The calculated magnitudes of the electric dipoles for $(TaSe_4)_2I$ surface (Fig. 5b) is around 0.098 eÅ per unit cell, comparable to the ferroelectric crystal $In_2Se_3$[53]. In Fig. 5c, the reaction path between the paraelectric and ferroelectric states of $(TaSe_4)_2I$ surface calculated by Nudged Elastic Band (NEB)[54,55] has been plotted, manifested by a small energy barrier (68 meV per surface I atoms), indicating that it is possible to overcome the energy barrier to access the less-favored ferroelectric states. However, for bulk $(TaSe_4)_2I$, the sliding-induced reconstruction of I atoms costs much energy, and in this case the symmetry in $z$-direction is protected to inhibit the occurrence of ferroelectricity, which is consistent with experimentally determined crystal structure of bulk $(TaSe_4)_2I$ crystals. Therefore, the origin of ferroelectricity can be attributed to the surface effect of I atom reconstruction rather than the intrinsic properties of bulk samples. Evidence of I atom shift has been found in the high-resolution scanning transmission electron microscope result as shown in Fig. S14. Especially, as a Weyl semimetal,[35,43] $(TaSe_4)_2I$ exhibits a gap opening due to the charge density wave (Fig. 5d–e), turning this system into an axion insulator, consistent with previous studies[35,39,43]. Thus, we also employed band structure calculations on the ferroelectric phase induced by the reconstruction of I atoms in $(TaSe_4)_2I$ surface *vs* the paraelectric phase. As shown in Fig. 5f–g, the nodal points at X and Y of the Brillouin zone occur in the paraelectric phase and open gaps in ferroelectric phase, indicating the interplay between ferroelectricity and band topology (The band structures of different $TaSe_4$ layers are calculated as plotted in Fig. S11, showing a good agreement). Note that here we only adopt an ideal cleaving of the crystal plane which 100% of the I atoms stay on surface. For the cleaved surface with reduced proportion residual iodine atoms, the surface nodal points tend to shift downward significantly[45] and may also correspond to different topological properties. The emergent ferroelectricity and band topology in $(TaSe_4)_2I$ makes it as a potential candidate for memristors, and may guide the design of ferroelectric topological devices.

In summary, we report the observation of room-temperature surface ferroelectric property in $(TaSe_4)_2I$ and further utilize it in prototype memristor devices. The ferroelectricity is evidenced by a hysteresis loop in PFM measurement. By employing the polarization-dependent Schottky barrier, we successfully realize the resistance switching of metal–$(TaSe_4)_2I$ contact with non-volatile property. The obtained $(TaSe_4)_2I$ memristor device shows a large on/off ratio up to $10^3$ with good endurance above $10^3$ times of switching. Theoretical calculation shows that the ferroelectricity in $(TaSe_4)_2I$ results from the surface reconstruction. It demonstrates $(TaSe_4)_2I$ as a new member of the on-demand low-dimensional ferroelectric materials[56,57] towards computation and energy conversion applications. The realization of ferroelectric memristors in a topological semimetal opens up new directions for novel topological electronic devices.

**Methods**

**Crystal growth and charactization**

The high-quality $(TaSe_4)_2I$ single crystals were prepared by standard chemical vapor transport method. Stoichiometric mixture of Ta, Se, and I was sealed in a 20 cm vacuum quartz tube with the ratio of 2: 7.6: 3.7. The tube was loaded in a two-zone furnace kept at 600 °C and 450 °C. After a 10-day growth procedure, shiny and needle-like single crystals were found in the low-temperature zone. Raman spectrum was obtained from a home-built system using 632.8 nm HeNe laser.

**Device fabrication and measurement**

The $(TaSe_4)_2I$ nanoribbons were mechanically exfoliated on silicon substrates with 285nm-$SiO_2$ by Scotch tape. Devices were fabricated by standard electron beam lithography with lift-off procedure with Cr/Ag/Au (5 nm/120 nm/120 nm) deposited as electrodes. The devices shown in the main text were annealed for 1 hour at 350 K in vacuum. Low-temperature electrical measurement was conducted in a commercial cryostat with Keithley 2450 Source Meter. The PFM test was carried out by applying a series of voltage pulses with the same duration while monitoring the electromechanical response of tip. No in-plane polarization was detected in $(TaSe_4)_2I$.

**DFT Calculations**

Density Functional Theory (DFT) calculations were performed using the projector-augmented wave method[58] as implemented in the Vienna Ab initio Simulation Package (VASP)[59,60]. The exchange-correlation interactions among electrons were described by the generalized-gradient approximation (GGA) with the Perdew–Burke–Ernzerhof revised for solids (PBEsol) functional[61]. The plane wave energy cutoff was set at 400 eV. All atomic positions are fully relaxed until the force acting on each atom became smaller than $1 \times 10^{-6}$ eV/Å. A vacuum layer of about 20 Å was adopted to avoid the interaction between periodic images. The Brillouin zone was sampled by a $5 \times 5 \times 1$ $k$-point mesh for the monolayer unit cell.

**Author contributions**
Cheng Zhang conceived the ideas and supervised the overall research. Xianghao Meng, Yuhan Du, and Xiang Yuan synthesized the crystals and performed optical characterization. Jianwen Ma, Yuxiang Wang, Yicheng Mou, Jiaming Gu, Jiayu Wang, and Wu Shi fabricated the devices and carried out the electric measurements. Bobo Tian, Chunsen Liu, Yannan Dai, Luqiu Chen, Xianghao Meng, and Haoqi Wu carried out the atomic force microscopy experiments. Wenting Lin and Lin Miao conducted the photoemission experiment. Haonan Wang and Zhenzhong Yang conducted the STEM experiments. Cheng Zhang and Jianwen Ma analyzed the data. Binhua Zhang, and Changsong Xu performed theoretical calculations. Cheng Zhang, Jianwen Ma, Xiang Yuan, Changsong Xu, Xianghao Meng, and Binhua Zhang wrote the paper with help from all other co-authors.

**Acknowledgments**

Cheng Zhang was sponsored by the National Key R&D Program of China (Grant No. 2022YFA1405700), the National Natural Science Foundation of China (Grant No. 12174069 and No. 92365104), and Shuguang Program from the Shanghai Education Development Foundation. Xiang Yuan was sponsored by the National Key R&D Program of China (Grant No. 2023YFA1407500), and the National Natural Science Foundation of China (Grants No. 12174104 and No. 62005079). Zhenzhong Yang acknowledged financial support from the National Key R&D Program of China (No. 2022YFA1402902). Part of the sample fabrication was performed at Fudan Nano-fabrication Laboratory. The authors gratefully thank Yi Cao, Du Xiang for helpful supports on the response time test.

**Competing financial interests**

The authors declare no competing financial interests.

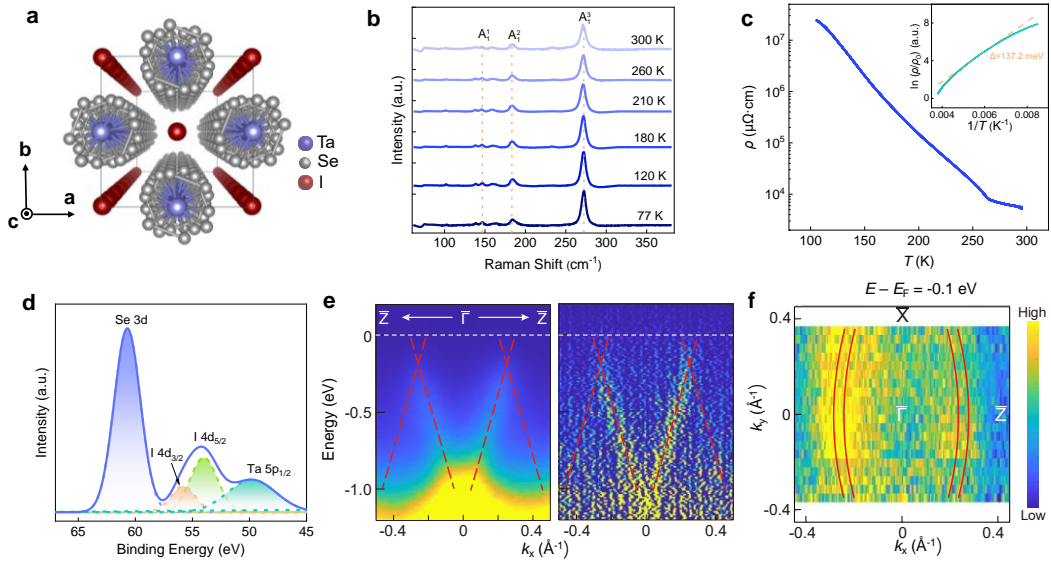

**Fig. 1.** (color online) The crystal structure and characterizations of (TaSe$_4$)$_2$I bulk crystals. (a) Top view of the unit cell of (TaSe$_4$)$_2$I. (b) Raman spectra of bulk (TaSe$_4$)$_2$I under different temperatures. (c) Longitudinal resistivity $\rho$ versus temperature $T$ (The inset shows the activation energy of 137.2 meV). (d) X-ray photoelectron spectra of bulk (TaSe$_4$)$_2$I. (e) ARPES band map of (TaSe$_4$)$_2$I along the $\overline{\Gamma Z}$ direction, the red dash lines are guidelines of band structure (left). The right panel is the second-order differential of the left panel. The incident photon energy is 21.2 eV. (f) CEC mapping acquired on the $\overline{X\Gamma Z}$ plane at $E-E_F$ = -0.1 eV.

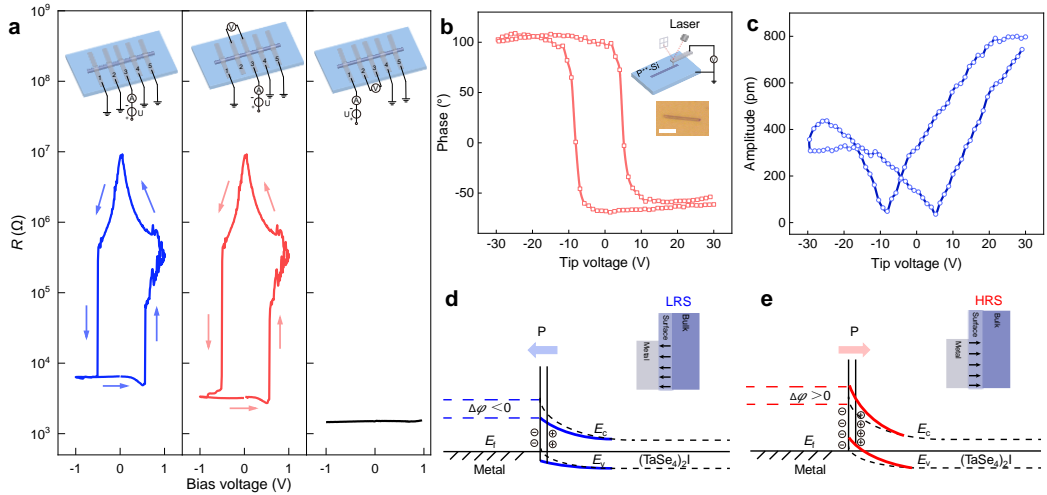

**Fig. 2.** (color online) Ferroelectric polarization reversal under external electrical field and resistance switching mechanism. (a) The relationship of resistance and voltage under different test configurations at 120 K including Two-terminal test (left panel), Interface test (middle panel), Four-terminal test (right panel). (b–c) PFM phase (b) and amplitude (c) hysteresis loop of a (TaSe$_4$)$_2$I thin nanoribbon on p$^{++}$-Si at room temperature. The inset shows the schematic of PFM setup and optical micrograph of the (TaSe$_4$)$_2$I flake. The scale bar is 5 μm. (d–e) Schematic illustration of resistance switching mechanism. The energy band at the interface of the contact is modified by the out-of-plane ferroelectric polarization. Upon applying the large negative voltage beyond the threshold, as shown

in (d), the surface layer produces upward ferroelectric polarization, driving the presence of LRS. Similarly, applying large positive voltage generates HRS, as shown in (e).

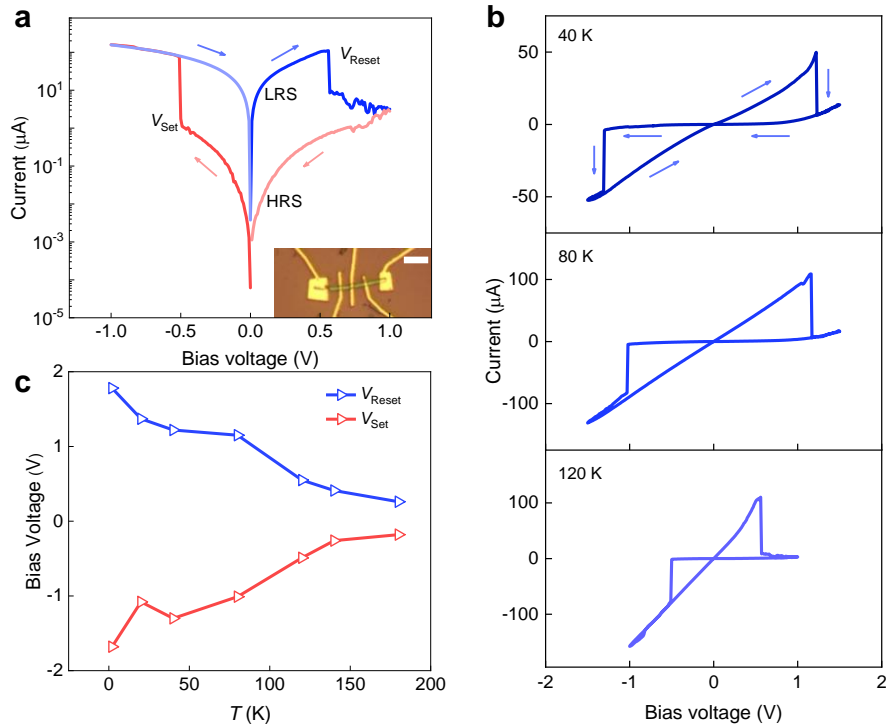

**Fig. 3.** (color online) *I–V* characteristics of the $(TaSe_4)_2I$ memristor. (a) Log scale *I–V* curves of the $(TaSe_4)_2I$ memristor at 120 K. The inset exhibits the optical micrograph of the $(TaSe_4)_2I$ device. The scale bar is 10 μm. (b) Linear scale *I–V* curves of the $(TaSe_4)_2I$ memristor at different temperatures. (c) Temperature dependence of set voltage and reset voltage.

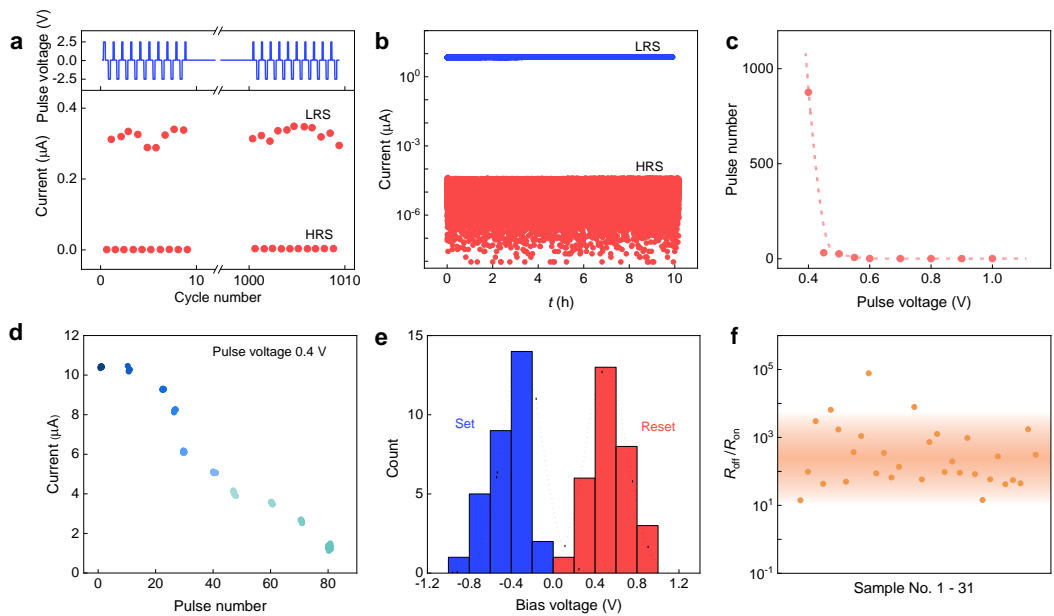

**Fig. 4.** (color online) The device performance of the $(TaSe_4)_2I$ memristor. (a) Endurance properties of the device. The HRS and LRS states of the device were switched by applying a series of ±2.5 V

pulse of 500 ms. Each voltage pulse is separated by time interval of 1 s. (b) Retention of the HRS and LRS currents. Multiple ±2.5 V voltage pulses of 500 ms were used to fully switch the device resistance states. (c) The required set number to reach the HRS with different magnitudes of the applied voltage. (d) Multi-level non-volatile current states as tuned by different pulse numbers, pulse voltage is 0.4 V. (e) Statistical distribution of the Set/Reset voltages for 31 devices. (f) The distribution of $R_{off}/R_{on}$ among different devices.

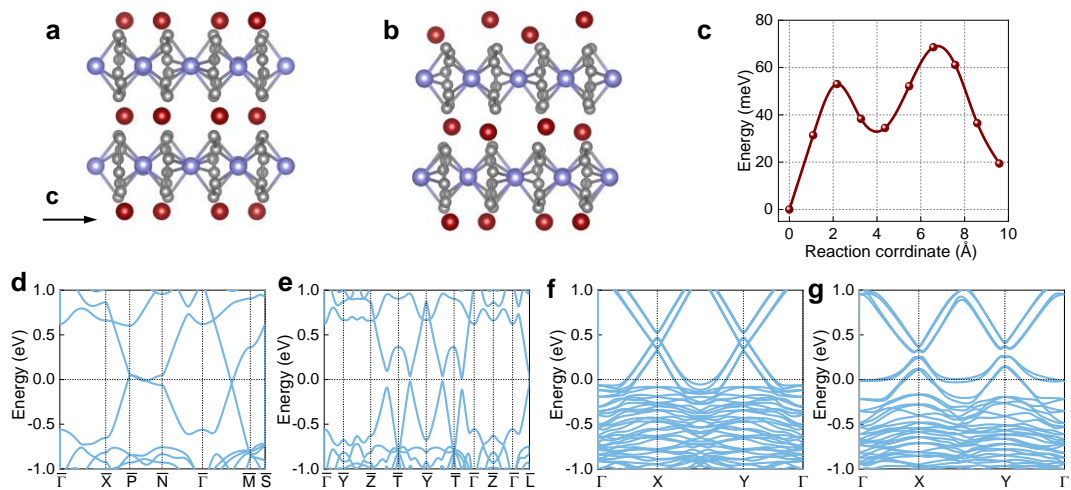

**Fig. 5.** (color online) Possible origin of ferroelectricity in $(TaSe_4)_2I$. (a–b) Side view of the $(TaSe_4)_2I$ (110) surface in (a) pristine paraelectric phase and (b) ferroelectric phase. Here, Iodine layer in the surface is dramatically buckling due to surface reconstruction. It effectively breaks the centrosymmetry and induces out-of-plane electric polarization. Purple, silver, and wine balls represent the Ta, Se, and I atoms, respectively. (c) The calculated NEB evolution of the relative total energy with respect to the pristine paraelectric phase in the two layers $(TaSe_4)_2I$ when the structure transforms from the phases with the paraelectric polarization in (a) to the electric polarization in (b). (d) The band structure of the undistorted pristine bulk $(TaSe_4)_2I$. (e) The band structure of the bulk $(TaSe_4)_2I$ with charge density wave distortion. (f–g) The band structure of $(TaSe_4)_2I$ surfaces in (a) and (b). Corresponding high-symmetric path in the Brillouin zone is plotted in Fig. S13.